\documentclass[a4paper,alpha-refs]{aejstyles}

\journal{aej}

\usepackage{graphicx}
\usepackage{siunitx}
\usepackage{eurosym}

\title{Per aspera ad astra simul: Through difficulties to the stars together}

\author[1,2\authfn{1}]{David Jones}
\author[3,\authfn{2}]{Petr Kab\'ath}
\author[1,2]{Jorge Garc\'ia-Rojas}
\author[4]{Josef Hanu\v s}
\author[5]{Marian Jakub\'ik}
\author[6]{Jan Jan\'ik}
\author[7]{Roman Nagy}
\author[7]{Juraj T\'oth}

\affil[1]{Instituto de Astrof\'isica de Canarias, E-38205 La Laguna, Spain}
\affil[2]{Departamento de Astrof\'isica, Universidad de La Laguna, E-38206 La Laguna, Spain}
\affil[3]{Astronomical Institute, Czech Academy of Sciences, Fri\v cova 298, 25165, Ond\v rejov, Czech Republic}
\affil[4]{Charles University, Faculty of Mathematics and Physics, Institute of Astronomy, V Hole\v sovi\v ck\'ach 2, CZ-18000, Prague 8, Czech Republic}
\affil[5]{Astronomical Institute, Slovak Academy of Sciences, 05960 Tatransk\'a Lomnica, Slovakia}
\affil[6]{Department of Theoretical Physics and Astrophysics, Masaryk University, Kotl\'a\v rsk\'a 2, CZ-611 37 Brno, Czech Republic}
\affil[7]{Faculty of Mathematics, Physics and Informatics, Comenius University, Bratislava, Slovakia}

\authnote{\authfn{1}djones@iac.es}
\authnote{\authfn{2}petr.kabath@asu.cas.cz}

\papercat{Resources \& Activities}

\runningauthor{Jones, Kab\'ath \& Garc\'ia-Rojas}

\jvolume{00}
\jnumber{0}
\jyear{2022}

\begin{document}

\begin{frontmatter}
\maketitle
\begin{abstract}

In this article, we detail the strategic partnerships ``Per Aspera Ad Astra Simul'' and ``European Collaborating Astronomers Project: Espa\~na-Czechia-Slovakia''.  These strategic partnerships were conceived to foment international collaboration for educational activities (aimed at all levels) as well as to support the development and growth of early career researchers.  The activities, carried out under the auspices of these strategic partnerships, demonstrate that Key Action 2 of the Erasmus+ programme can be an extremely valuable resource for supporting international educational projects, as well as the great impact that such projects can have on the general public and on the continued development of early career researchers. We strongly encourage other educators to make use of the opportunities offered by the Erasmus+ scheme.
\end{abstract}


\end{frontmatter}


\section{Introduction}

Many will be familiar with the Erasmus+ programme of the European Commission through its scheme to support students spending a semester/year abroad during their undergraduate degrees (generally as part of its Key Action 1).  However, Key Action 2 of the Erasmus+ programme provides funding for multinational strategic partnerships between institutions with the aim of promoting educational, professional and personal development.  These strategic partnerships are applied for and managed on the national level with, for example, the Czech national agency having budgets of approximately 1.5~M\euro{} and 3~M\euro{} in the 2017 and 2020 calls, respectively. \href{http://erasmus.asu.cas.cz/erasmus2017/index.html}{``Per Aspera Ad Astra Simul''} (2017-2020, 2017-1-CZ01-KA203-035562) and its (currently active) successor \href{http://erasmus.asu.cas.cz/index.html}{``European Collaborating Astronomers Project: Espa\~na-Czechia-Slovakia''} (2020-2023, 2020-1-CZ01-KA203-078200), the subjects of this article, are excellent examples of how this scheme can support astronomy education and the professional development of early career researchers (ECRs).

\section{Background}
\label{sec:background}

The Erasmus+ program is designed to support education, training and youth through international cooperation. In such a competitive environment, the development of ECRs in astronomy rely heavily on the most modern techniques (be they observational or theoretical) and international collaboration. Therefore, it is crucial for young astronomers to acquire as much wide-ranging expertise and international experience as possible during their studies. Ultimately, this makes ECRs better scientists with improved career prospects.

Although Spain and the Czech Republic joined the European Southern Observatory (ESO -- Europe's premier ground-based astronomy consortium) almost simultaneously\footnote{At the time of publication Slovakia is not an ESO member state.} (late 2006/early 2007), the Spanish astronomical community is objectively larger and more developed \citep{barcons07,palous07}. Spain has contributed more staff to the ESO observatories (indeed, the current Director General is Spanish), won more industrial ESO contracts, and been awarded more observing time on ESO facilities. Spain is also home to several major astronomical facilities, including the World's largest optical-infrared telescope -- the Gran Telescopio Canarias (GTC).

The Czech and Slovak astronomical communities are smaller, but with a rich history particularly in the fields of theoretical astrophysics, variable stars and minor solar system bodies.  More recently, both communities have expanded significantly in the theoretical and observational study of exoplanets.  Several 1--2m class telescopes have been built in Czechia and Slovakia, but the continental weather impacts significantly upon their usage. 

The Erasmus+ strategic partnerships ``Per Aspera Ad Astra Simul'' and ``European Collaborating Astronomer Projects: Espa\~na-Czechia-Slovakia'' were conceived as mutually beneficial agreements between institutes in Spain, the Czech Republic and Slovakia. The significant Spanish expertise in large observing facilities offers a clear opportunity for learning and growth for ECRs from the Czech and Slovak partners.  Similarly, Spanish ECRs would greatly benefit from extended research stays at the Czech and Slovak partners, developing long-lasting collaborations and gaining valuable international experience. Furthermore, the Erasmus+ partnerships fosters collaboration also among senior staff, ensuring continued benefit to further generations of ECRs beyond the duration of the project.

\section{Partners and their roles}
\label{sec:partners}

The partners of the aforementioned Erasmus+ strategic partnerships are listed in Table \ref{tab:partners}.  The lead institute of the project is the Astronomical Institute of the Academy of Sciences of the Czech Republic (AI ASCR), the largest astronomical institute in the Czech Republic.  The Department of theoretical physics and astrophysics of Masaryk University and the Astronomical Institute of Charles University (AI CU, who joined in 2020) are the other Czech partners in the project and comprise two of the longest established and important astronomical institutes in the Czech Republic.

In Slovakia, the Astronomical Institute of the Slovak Academy of Sciences (AI SAS) and Comenius University Bratislava are both partners.  Spain is represented in the partnership by the Instituto de Astrof\'isica de Canarias (IAC), the largest astronomical institute in the country and the operators of the Teide and Roque de los Muchachos Observatories, the latter of which is home to the GTC.  Indeed, GranTeCan (the public company responsible for the GTC's design, construction, and continued operation) was an associate partner from 2017--2020.

As lead institute, AI ASCR has been responsible for the administration and coordination of the project and the associated funds.  AI SAS and Masaryk University have organised summer schools for ECRs in 2019 and 2020, respectively.  Further summer schools, organised by the IAC and Comenius University Bratislava are planned for the coming years.  All partners have participated in mobilities of ECRs and more senior researchers (see Section \ref{sec:mobilities}), and all have been involved in local outreach and educational activities (see Section \ref{sec:outreach}).  Similarly, senior researchers from all partner institutions contributed to a book of thirteen detailed reviews, designed to be useful graduate-level introductions to the topics ranging from meteors and space debris through to the formation of galaxies and the standard cosmological model, and which was published in 2020 \citep{kabath20}.

\begin{table*}[bt!]
\caption{List of partner institutions}\label{tab:partners}
\begin{tabularx}{\linewidth}{l l L L}
\toprule
{Institution} & {Country} & {Notes}\\
\midrule
Astronomical Institute of the Academy of Sciences of the Czech Republic (AI ASCR) & Cz & Lead institute\\
Instituto de Astrof\'isica de Canarias (IAC) & Es & Summer school organiser 2022\\
Comenius University Bratislava & Sk & Summer school organiser 2023\\
Masaryk University & Cz & Summer school organiser 2020\\
Astronomical Institute of the Slovak Academy of Sciences (AI SAS) & Sk & Summer school organiser 2019\\
Astronomical Institute of Charles University (AI CU) & Cz  & Partner since 2020\\
GranTeCan & Es &  Associate partner until 2020\\
\bottomrule
\end{tabularx}
\end{table*}

\section{Mobilities}
\label{sec:mobilities}

The strategic partnerships offered significant financial support for mobilities of researchers between the partner institutions.  Short-term ($\sim$a few weeks) exchanges of more senior researchers designed to foment collaborations between the institutions which can then be used to achieve the objectives of the Erasmus+ programme\footnote{For further details of these objectives, please see \href{https://erasmus-plus.ec.europa.eu/programme-guide/part-a/priorities-of-the-erasmus-programme/objectives-features}{What are the objectives of the Erasmus+ programme?}}, as well as longer-term (2--6 months) exchanges of ECRs with the aim of aiding their personal and professional development.  At the end of the stay, the researchers were encouraged to summarise their activities in the form of a blog post on the project's \href{http://erasmus.asu.cas.cz/erasmus2017/blog.html}{webpages}.

In total more than 30 international exchanges have already been completed (in spite of the obvious problems posed by the COVID-19 pandemic striking part way through the execution of the first strategic partnership), with further exchanges planned for 2022 and 2023.  In many cases, the benefits of these exchanges have been tangible and obvious \citep[for example, resulting in peer-reviewed publications in high impact journals;][]{jones19,gonzalez20,paunzen21}, while in other cases the benefits are more ``soft'' -- leading to long-term collaborations, personal and professional development, improved career prospects, etc. 

\subsection{Gran Telescopio Canarias}
\label{sec:tereza}

A special case among the ECR exchanges, where the majority were conceived as collaborations on blue skies astronomy research under the supervision of a senior researcher at the receiving institution, was the long-term mobility of one ECR to GranTeCan.  During their stay, the PhD student combined continued research towards their doctoral thesis with support astronomer activities at the World's largest optical-infrared telescope -- the 10.4-m Gran Telescopio Canarias (GTC).  This offered the student a unique opportunity to obtain in-depth experience of the tasks involved in the day-to-day operations of a world-class astronomical observatory -- experience which has undoubtedly been extremely valuable in their career development, with the student going on to take up highly competitive positions at the European Space Agency (ESA) and European Southern Observatory (ESO) following the completion of their mobility and thesis defense.

\section{Summer schools}
\label{sec:schools}

In June 2019, the summer school \href{https://opticon-schools.nbi.ku.dk/other-schools/from-proposals-to-publication/}{``Observational astrophysics: from proposals to publication''} was organised in Tatransk\'a Lomnica (Slovakia) under the auspices of the Per Aspera Ad Astra Simul grant and in collaboration with OPTICON.  The school comprised two parts: hands-on projects with archival data and lectures/activities to teach the participants about observing time proposal evaluation procedures.

The hands-on projects were undertaken in groups of approximately four students (with a total of approximately 40 students in attendance) under the supervision of an experienced tutor.  The projects themselves covered topics ranging from near-Earth asteroids \citep{NEAs} through to binary quasars \citep{quasars} all based on publicly available archive data.  Upon completion of these projects, the participating students presented their results in the form of a mini-conference and later reported their analyses and results in papers published in the Contributions of the Astronomical Observatory Skalnat\'e Pleso journal \citep{summerschool}.

A second summer school, \href{https://gate.physics.muni.cz/}{``GAIA and TESS: Tools for understanding the Local Universe''}, was organised in 2020, originally to be hosted in Brno (Czech Republic) but ultimately held remotely due to the COVID-19 pandemic (with 17 students attending via video conference).  The school featured talks on the capabilities and applications of the Gaia and TESS missions by highly experienced experts, as well as hands-on sessions and research projects making use of the public data products of these missions \citep{2021CoSka..51...41S}.

A third school, \href{https://iacerasmus.github.io/ERASMUS2022/}{``Eclipsing Binaries and Asteroseismology: Precise fundamental stellar parameters in the golden age of time-domain astronomy"}, will be hosted on the Spanish island of La Palma in September of 2022.  This school will be run in a hybrid format with 15 in-person students and up to 200 attending via video conference.  A final school is planned to be held in Bratislava (Slovakia) in 2023.

\section{Outreach}
\label{sec:outreach}

Thus far, we have discussed the activities of the strategic partnerships targeted towards the education and professional development of ECRs.  However, the project has always had a significant component geared towards the education of a younger and/or wider audience, principally in the form of activities in local schools and public talks.

We stress the importance of early education and thus target our effort to work together with the local schools and preschool classes.  In Ond\v{r}ejov, activities were focused towards children of age 5--6 years \footnote{Similar activities for older children were planned, but have been significantly impacted by the COVID-19 pandemic.}, preparing for school which begins at the age of 6 in the Czech Republic. We invite the preschoolers to observatory and to the 2-m telescope dome, where they can see the telescope. In the dome, we present a short talk about a selected topic, ranging from the solar system to the life of stars, which the children will then later build upon in class with their teachers.  After the visit, the children prepare art work or a similar project inspired by the selected topic, which is then presented in a small gathering with their parents invited. These activities not only engage the children with their first experience of space and science, but also help them to develop creative and presentation skills. 
In Spain, activities were organised for local 4th grade students (ages 9--10) which focused on the importance of astronomy in the Canary islands throughout history, from the ancient aborigines through to the current day.  These included hands-on workshops on the solar system and the timescales of the Universe, offering many students their first contact with a professional astronomer. Activities for high school students were also organized, focusing on different aspects of technological development related to astronomy and on a general view of the Universe through the different ranges of the electromagnetic spectrum.  Similarly, members of our team contributed to the long-running ``Nuestros Alumnos en el Roque de Los Muchachos'' programme, giving guided tours of the Roque de Los Muchachos Observatory (including going inside the GTC dome) to local high school students in the 4th grade of compulsory secondary education (4$^o$ ESO, age 15).

For the general public, several popular lectures were delivered by members of the team in planetariums and museums across partner countries.  Talks for citizen scientists and amateur astronomers were also organised, in particular outlining the potential for observations of stellar occultations by asteroids.     Similarly, a \href{https://www.youtube.com/channel/UCXBLE1tzrL2mhY3Kb0ELieQ/videos}{YouTube channel} was created for the project, hosting educational videos on the history of astronomy, astronomical techniques and principles, and interviews.  Furthermore, the partners at Comenius University in Bratislava contributed to their pre-existing \href{https://www.youtube.com/playlist?list=PLqiGU4u5LkCF2YYU450gssOPPhE-4jOfI}{YouTube channel} as part of the project.  These open talks and YouTube videos serve not only to educate but also to strengthen the connection between the public and the scientists whom are at least partially supported by public funding.

\section{Conclusions}
\label{sec:conclusions}

Key Action 2 of the European Union's Erasmus+ programme offers substantial grant funding for international collaboration on educational projects.  The (on-going) outputs of two such grants, awarded to a consortium of Czech, Slovak and Spanish research institutes and universities, have been briefly presented here.  These grants have supported educational activities ranging from outreach events for young children through to summer schools for graduate students, and from public lectures and YouTube videos through to financial support for extended international research visits to support the professional development of ECRs.

The impact of the projects, particularly for the professional development of ECRs, has been clear.  A number of summer school participants have now completed their Masters or PhDs (stating that the skills acquired as part of the schools were useful for their research), while many of the ECRs to have research trips funded by the projects have since gone on to obtain prestigious fellowships (e.g.\ ESA) or permanent research/teaching positions (e.g.\ through the Mexican ``C\`atedra CONACYT para j\'ovenes investigadores'' scheme).

The work undertaken as part of ``Per Aspera Ad Astra Simul" and ``European Collaborating Astronomers Project: Espa\~na-Czechia-Slovakia" clearly demonstrates the potential impact that Erasmus+ Key Action 2 funding can facilitate.  We strongly encourage other astronomy educators to consider how they might make similarly good use of the available funding in future calls.






\section{Declarations}

\subsection{List of abbreviations}
\begin{itemize}
\item[AI ASCR] Astronomical Institute of the Academy of Sciences of the Czech Republic
\item[AI CU] Astronomical Institute of Charles University
\item[AI SAS] Astronomical Institute of the Slovak Academy of Sciences
\item[ECR] Early career researchers
\item[ESA] European Space Agency
\item[ESO] European Southern Observatory
\item[GTC] Gran Telescopio Canarias
\end{itemize}

\subsection{Ethical Approval}
Not applicable.

\subsection{Consent for publication}

Not applicable.

\subsection{Competing Interests}

The author(s) declare that they have no competing interests.

\subsection{Funding}

This work was supported by the Erasmus+ programme of the European Union under grant numbers 2017-1-CZ01-KA203-035562 and 2020-1-CZ01-KA203-078200 (PI P.\ Kab\'ath).


\subsection{Author's Contributions}

David Jones - Conceptualization, Funding acquisition, Project administration, Writing – original draft, Writing – review \& editing\\
Petr Kab\'ath - Conceptualization, Funding acquisition, Project administration, Writing – review \& editing, Project administration
Jorge Garc\'ia-Rojas - Project administration, Writing - review \& editing
Josef Hanu\v s - Project administration, Writing - review \& editing
Marian Jakub\'ik - Project administration, Writing - review \& editing
Jan Jan\'ik - Project administration, Writing - review \& editing
Roman Nagy - Project administration, Writing - review \& editing
Juraj T\'oth - Project administration, Writing - review \& editing


\section{Acknowledgements}

We would like to acknowledge the role of the researchers at the participating institutions without whom these partnerships would not have been a success -- the ECRs for their eagerness to participate in mobilities and schools, and senior researchers for their willingness to host ECR mobilities and to contribute to \citet{kabath20}.



\bibliography{paper-refs}

\begin{thebibliography}{}

\bibitem[{Altamura} et~al., 2019]{quasars}
{Altamura}, E., {Brennan}, S., {Le{\'s}niewska}, A., {Pint{\.{e}}r}, V., {dos
  Reis}, S.~N., {Geier}, S., and {Fynbo}, J.~P.~U. (2019).
\newblock {Discovery of a binary quasar at z = 1.76}.
\newblock {\em Contributions of the Astronomical Observatory Skalnate Pleso},
  49(3):528--531.

\bibitem[{Barcons}, 2007]{barcons07}
{Barcons}, X. (2007).
\newblock {Astronomy in Spain}.
\newblock {\em The Messenger}, 127:4.

\bibitem[{Gonz{\'a}lez Manrique} et~al., 2020]{gonzalez20}
{Gonz{\'a}lez Manrique}, S.~J., {Kuckein}, C., {Pastor Yabar}, A., {Diercke},
  A., {Collados}, M., {G{\"o}m{\"o}ry}, P., {Zhong}, S., {Hou}, Y., and
  {Denker}, C. (2020).
\newblock {Tracking Downflows from the Chromosphere to the Photosphere in a
  Solar Arch Filament System}.
\newblock {\em ApJ}, 890(1):82.

\bibitem[{Jones} et~al., 2019]{jones19}
{Jones}, D., {Pejcha}, O., and {Corradi}, R. L.~M. (2019).
\newblock {On the triple-star origin of the planetary nebula Sh 2-71}.
\newblock {\em MNRAS}, 489(2):2195--2203.

\bibitem[{Kab{\'a}th} et~al., 2020]{kabath20}
{Kab{\'a}th}, P., {Jones}, D., and {Skarka}, M. (2020).
\newblock {\em {Reviews in Frontiers of Modern Astrophysics; From Space Debris
  to Cosmology}}.
\newblock {Springer}.

\bibitem[{Kab{\'a}th} et~al., 2019]{summerschool}
{Kab{\'a}th}, P., {Korhonen}, H., and {Jones}, D. (2019).
\newblock {Observational astrophysics: from proposals to publication}.
\newblock {\em Contributions of the Astronomical Observatory Skalnate Pleso},
  49(3):522--527.

\bibitem[{Ku{\v{c}}{\'a}kov{\'a}} et~al., 2019]{NEAs}
{Ku{\v{c}}{\'a}kov{\'a}}, H., {Mikhalchenko}, O., {Popescu}, M., {Ransome}, C.,
  and {Sharma}, A. (2019).
\newblock {Optical spectra of near-Earth asteroids (381906) 2010 CL19 and
  (453778) 2011 JK}.
\newblock {\em Contributions of the Astronomical Observatory Skalnate Pleso},
  49(3):532--538.

\bibitem[{Palou{\v{s}}} and {Hadrava}, 2007]{palous07}
{Palou{\v{s}}}, J. and {Hadrava}, P. (2007).
\newblock {Astronomy in the Czech Republic}.
\newblock {\em The Messenger}, 128:3.

\bibitem[{Paunzen} et~al., 2021]{paunzen21}
{Paunzen}, E., {H{\"u}mmerich}, S., {Fedurco}, M., {Bernhard}, K.,
  {Kom{\v{z}}{\'\i}k}, R., and {Va{\v{n}}ko}, M. (2021).
\newblock {V680 Mon - a young mercury-manganese star in an eclipsing heartbeat
  system}.
\newblock {\em MNRAS}, 504(3):3749--3757.

\bibitem[{Skarka} et~al., 2021]{2021CoSka..51...41S}
{Skarka}, M., {Jan{\'\i}k}, J., {Paunzen}, E., and {Glos}, V. (2021).
\newblock {The GATE summer school}.
\newblock {\em Contributions of the Astronomical Observatory Skalnate Pleso},
  51(1):41--44.

\end{thebibliography}

\end{document}